\documentclass[11pt]{article}
\usepackage{graphicx}
\newcommand{\be}{\begin{equation}}
\newcommand{\ee}{\end{equation}}
\newcommand{\bea}{\begin{eqnarray}}
\newcommand{\eea}{\end{eqnarray}}
\newcommand{\bean}{\begin{eqnarray*}}
\newcommand{\eean}{\end{eqnarray*}}
\begin{document}
\setlength{\hoffset}{-75pt}
\setlength{\marginparsep}{1pt}

\title{\bf A simple background-independent \\ hamiltonian 
quantum model} 
\author{Daniele Colosi and Carlo Rovelli\\[2mm] \em Centre de
Physique Th\'eorique de Luminy, CNRS, F-13288, France.\\
\em Dipartimento di Fisica Universit\`a di Roma ``La Sapienza", Roma,
Italia.} \date{\small\em\today} 

\maketitle

\begin{abstract} 
We study formulation and probabilistic interpretation of a simple
general-relativistic hamiltonian quantum system.  The system has no
unitary evolution in background time.  The quantum theory yields
transition probabilities between measurable quantities (partial
observables).  These converge to the classical predictions in the
$\hbar\to 0$ limit.  Our main tool is the kernel of the projector on
the solutions of Wheeler-deWitt equation, which we analyze in detail. 
It is a real quantity, which can be seen as a propagator that
propagates ``forward" as well as ``backward" in a local parameter
time.  Individual quantum states, on the other hand, may contain only
``forward propagating" components.  The analysis sheds some light on
the interpretation of background independent transition amplitudes in
quantum gravity.
\end{abstract}

\vskip.5cm

\section{Introduction}

Conventional mechanics describes evolution in ``background time": the
independent time variable $t$ is assumed to represent a measurable but
non-dynamical physical quantity.  This is true in the classical as
well as in the quantum theory, where evolution in background time is
given by unitary transformations.  With general relativity, however,
we have learned that there is no non-dynamical measurable time in
nature: there is no background spacetime, and no background time in
particular.  Therefore, at the fundamental level physics cannot
describe evolution in time.  It can only describe relations, or
correlations, between measurable quantities.  It is therefore
necessary to extend the formalism of classical and quantum mechanics
to such a background independent context.

A strategy in this direction has been explored in a number of recent
works \cite{partial,2,book}.\footnote{For alternative systematic
attempts to formulate general covariant quantum theory, see
\cite{hartle,halliwell}.} Here we study a simple system in order to
illustrate and test ideas and techniques discussed in these works.  We
do not restate the general theory here, for which we refer the reader
to the references above, but this paper is nevertheless
self-contained.  The system we consider has one degree of freedom.  It
has a well defined classical dynamics.  It could describe a simple
cosmological model -- or a simple mechanical system studied without
using an external clock.  The system has two variables, which we call
$a$ and $b$.  Locally, we can view $b$ as the independent variable and
$a$ as the dependent one.  The model describes then the evolution
$a(b)$ of $a$ as a function of $b$.  In other words, locally we can
view $b$ as the time and $a$ as the physical degree of freedom.  But
globally this picture breaks down: the dynamics predicts relations
betweent the two variables $a$ and $b$ which are on an equal footing. 
The quantum evolution in $b$ fails to be unitary.  This is the
difficulty that generates the much debated ``problem of time" in
quantum general relativity \cite{isham,time}.

The system we consider has been investigated in \cite{halliwell,cr}. 
The difficulty is not to define the quantum states of the theory -- a
rather easy task using standard hamiltonian methods.  The difficulty
is to define a consistent probabilistic interpretation of the
formalism.  Due to the lack of unitary time evolution, the
conventional probabilistic interpretation of the wave function at
fixed time does not apply.  A good discussion of the difficulties of
the ``naive" interpretation of the wave function as a probability
density, and of tentative strategies to solve the problem can be found
in \cite{halliwell}.  In \cite{cr}, one of us has studied the physical
interpretation of this system using operators well defined on the
physical Hilbert space.  These describe ``complete" observables, (or
``perennials") and include, in particular the ``evolving constants of
the motion" \cite{time,cr}.  This way of interpreting general
relativistic quantum systems is correct in principle, but may be
cumbersome in practice.

In \cite{partial}, it was pointed out that the conventional notion of
``observable" may not be the most suitable one in a background
independent context.  The weaker notion of ``partial observable" may be
more useful in such a context.  A partial observable is any quantity
that can be measured -- even if it cannot be predicted by the theory. 
Using this notion, general relativistic systems admit a simple
reading: they predict correlations between partial observables.  This
is the point of view developed in \cite{partial,2,book}.  In this
philosophy the key object that yields the interpretation of the
quantum formalism is the ``propagator" $K(a,b,a',b')$.  This is
defined as the kernel of the projector from the kinematical Hilbert
space $\cal K$ (the space representing all possible outcomes of
measurements of partial observables) to the physical Hilbert space
$\cal H$ (the space representing the sole outcomes of measurements of
partial observables that are allowed by dynamics).  The propagator
plays a role similar to the propagator of the Schr\"odinger equation
in nonrelativistic quantum mechanics, which, as emphasized by Feynman,
expresses the quantum dynamics.  The propagator $K(a,b,a',b')$ can be
given a direct physical interpretation \cite{2}.  This point of view
might prove useful for the interpretation of the formalisms of non
perturbative quantum gravity -- in particular in loop quantum gravity
\cite{loop} and its sum over path version, the spinfoam formalism
\cite{spinfoam}, which gives us the analog of $K(a,b,a',b')$ in
quantum general relativity \cite{tutti}. 

Here we study the propagator $K(a,b,a',b')$ of our simple model and we
discuss physical interpretation and classical limit.  We show that in
a regime in which $b$ can be taken as an independent time parameter
the probabilistic interpretation proposed in \cite{2} reduces to the
conventional interpretation of the modulus square of the wave function
as a spacial probability density at fixed time.  We illustrate the
semiclassical limit and some semiclassical states.  The propagator
turns out to be real.  It can be seen as propagating ``forward" as
well as ``backward", in a local parameter time, in spite of the fact
that the difference between forward and backward propagation is not
manifest in the classical theory.  The quantum states, on the other
hand, split among forward and backward propagating ones.  This might
be the same situation that we find in quantum general relativity.

In the section 2 we discuss the classical theory.  We formulate the
theory in a fully covariant language, that illustrate the structure of
general-relativistic hamiltonian mechanics.  We study the quantum
theory in Section 3.  In the appendices, we recover an approximate 
Schr\"odinger equation, we recall the probabilistic interpretation
based on the flux of the conserved current, and we illustrate a
derivation of the propagator using functional integral techniques.  A
complete formulation and a general discussion of the classical and
quantum formalism that we utilize here can be found in chapters 3 and
5 of reference \cite{book}.

\section{Classical theory}

\subsection{Definition of the model}

A simple harmonic oscillator can be viewed as system with two partial
observables, $q$ and $t$.  A motion of the system defines a relation
between $q$ and $t$.  A given motion is characterized by the two
constants $A\in[0,\infty]$ and $\phi\in[0,2\pi]$, and is given by the
equation
\be f(q,t)=q-A\sin(\omega t+\phi)=0, \ee
where $\omega$ is a constant characterizing the system.  This motion
is a sinusoid of amplitude $A$ and phase $\phi$ in the $(q,t)$ plane,
the extended configurations space $\cal C$.  This dynamics is the
classical limit of a quantum dynamics.

Here, we consider a system with two partial observables $a$ and $b$. 
A given motion of the system is characterized by two constants
$A\in[0,\sqrt{2M}]$ and $\phi\in[0,2\pi]$, and is given by the
equation
  \be f(a,b)= \left(\frac{a}{A}\right)^2+ \left(\frac{b}{B}\right)^2 -
  2\cos\phi\ \frac{a}{A}\frac{b}{B}=\sin^2\phi,
  \label{ellipses}
\ee
where $M$ is a constant characterizing the system and $B^2\equiv
2M-A^2$.  This motion is an ellipse of radii $A$ and $B$ and
inclination $\phi$ in the $(a,b)$ plane, the extended configuration
space $\cal C$.  We want to understand if this system can be seen as
the classical limit of a quantum system, with a well defined
probabilistic interpretation.

\subsection{Classical dynamics}

A relativistic hamiltonian formulation of the system above can be
given on the extended configuration space ${\cal C}=R^2$ coordinatized
by the partial observables $(a,b)$. 
The dynamics (\ref{ellipses}) is governed by the relativistic
hamiltonian 
\bea 
H(a,b,p_a,p_b)=\frac{1}{2} \left( a^2+b^2+p_a^2+p_b^2 \right) - M
\label{H}
\eea
defined on the cotangent space $T^*{\cal C}=R^4$ coordinatized by
$(a,b,p_{a},p_{b})$, where $p_{a}$ and $p_{b}$ are the momenta
conjugate to $a$ and $b$.  In general, in a general-relativistic
hamiltonian system the dynamics is defined by the surface
$H(q^a,p_a)=0$, in $T^*{\cal C}$ (see chapter 3 of \cite{book}.)  The
hamiltonian (\ref{H}) yields the Hamilton-Jacobi equation
\bea 
\left(\frac{\partial S(a,b)}{\partial a}\right)^2
+\left(\frac{\partial S(a,b)}{\partial b}\right)^2 + a^2+b^2- 2M = 0.
\label{HJ}
\eea
Given a one parameter family $S(a,b;A)$ of solutions of this equation,
the physical motions are determined --given two constants $A$ and
$p_{A}$-- by the equation
\bea 
\frac{\partial S(a,b;A)}{\partial A}-p_{A} = 0.
\label{solut}
\eea
A family of solutions of the Hamilton-Jacobi equation (\ref{HJ}) is 
given by 
\bea 
S(a,b;A) = 
 \frac{a}{2} \sqrt{A^2 -a^2} + 
\frac{A^2}{2} \arctan \left( \frac{a}{\sqrt{A^2 -a^2}} 
\right) +\frac{b}{2} \sqrt{B^2 -{b}^2} + \frac{B^2}{2} 
\arctan \left( \frac{b}{\sqrt{B^2 -{b}^2}} \right).
\label{eq:HJfunction}
\eea
inserting this in (\ref{solut}) gives (\ref{ellipses}), where
$p_{A}=A\phi$, with simple algebra.

\subsection{Action}

Consider a path $\lambda:[0,1]\to T^*{\cal C}$ in the space of the
coordinates $a,b$ and their momenta $p_{a},p_{b}$, and let
$\tilde\lambda:[0,1]\to {\cal C}$ be its projection on the extended
configuration space $\cal C$ (just the coordinates $a,b$).  The action
$S[\lambda]$ is defined by
\bea 
\label{action}
S[\lambda] =  \int_{\lambda}\ (p_{a}da+p_{b}db).  
\eea
The physical motions $\tilde\lambda$ are the ones such that $\lambda$
extremizes this action in the class of paths that satisfy 
\bea 
H(a,b,p_{a},p_{b}) = 0 
\label{Huz}
\eea
and such that $\tilde\lambda$ has fixed extrema
$\tilde\lambda(0)=(a,b)$ and $\tilde\lambda(1)=(a',b')$.  These
motions follow the ellipse (\ref{ellipses}).  This is a very general
structure, common to all non relativistic as well as general
relativistic systems \cite{book}.

\subsection{Geometry}

Since the space $T^*{\cal C}$ is a cotangent space, it is naturally
equipped with the Poincar\'e one-form
\bea 
\theta = p_{a} da + p_{b} db 
\label{theta}
\eea
and the symplectic two-form $\omega=d\theta$.  Notice that the action
(\ref{action}) is simply the line integral along the path of the
Poincar\'e one-form
\bea 
\label{action2}
S[\lambda] =  \int_{\lambda} \theta.
\eea
Equation (\ref{Huz}) defines a surface $\Sigma$ in the space $T^*{\cal
C}$.  In the model we are considering this surface is a three-sphere
of radius $\sqrt{2M}$.  The restriction of $\omega$ to $\Sigma$ is
degenerate: it has a null direction $X$.  That is, there is a vector
field $X$ on $\Sigma$ that satisfies $\omega(X)=0$.  It can be easily
verified that this is
\bea 
X=
p_{a}\ \frac{\partial}{\partial a} +
p_{b}\ \frac{\partial}{\partial b} -
a\ \frac{\partial}{\partial p_{a}} -
b\ \frac{\partial}{\partial p_{b}}.
\label{X}
\eea
The integral lines of $X$ on $\Sigma$ define the motions.  It is easy
to see that these integral lines are circles.  In fact, they define a
standard Hopf fibration of the three-sphere $S_{3}\sim S_{2}\times
S_{1}$.  The motions are precisely the ones given by (\ref{ellipses}),
and the corresponding momenta are
\bea \label{eq:momenta}
p_a &=&  \sqrt{A^2 - a^2},  \nonumber  \\
p_b &=& \sqrt{B^2 - b^2}.  
\eea
The quotient of $\Sigma$ by the motions  $S_{2}\sim S_{3}/
S_{1}$ is the physical phase space of the system, coordinatized by 
the two constants $A$ and $\phi$
\bea
2A^2 &=& 2M+p_a^2-p_b^2 +a^2-b^2 \\
\tan \phi &=& \frac{p_a b - p_b a}{p_a p_b + ab},
\eea
which are constant along the motions.

\subsection{Parametrization and the associate nonrelativistic
system}

The dynamics of the system becomes easier to deal with if we introduce
a non-physical parameter $\tau$ along the physical motions
(\ref{ellipses}).  We can parametrize the ellipse (\ref{ellipses}) 
as 
follows 
\bea 
\label{eq:coord} 
a(\tau)&=& A\ \sin(\tau +\phi), \nonumber \\ 
b(\tau) &=& B\ \sin(\tau ). 
\eea
The momenta are then given by  
\bea \label{eq:momenta2} p_a(\tau) &=& \sqrt{A^2 - a^2}\ =\ A\
\cos(\tau +\phi)\ =\ \frac{da(\tau)}{d\tau}, \nonumber \\
p_b(\tau) &=& \sqrt{B^2-b^2}\ =\ B\ \cos(\tau)\ =\
\frac{db(\tau)}{d\tau}.  \eea
The parameter $\tau$ is not connected with observability; it is only
introduced for convenience.  In particular the physical predictions of
our system regard the relation between $a$ and $b$, not the dependence
of $a$ and $b$ on $\tau$.  The parameter $\tau$ corresponds to the
time coordinate of general relativity: physically meaningful
quantities are independent from the time coordinate.

It is useful to consider also a \emph{distinct} dynamical system with
\emph{two} degrees of freedom $a(\tau)$ and $b(\tau)$ evolving
according to equations (\ref{eq:coord}) in a time parameter $\tau$. 
We call this system the \emph{associate} nonrelativistic system.  It
is important not to confuse the two systems.  The associate system has
two degrees of freedom, while our original general-relativistic system
has one degree of freedom.  The associate system is governed by the
function $H$ given in (\ref{H}), seen now as a conventional
non-general-relativistic hamiltonian.  Indeed, it is immediate to see
that equations (\ref{eq:coord}-\ref{eq:momenta2}) are the solutions of
the Hamilton equations of $H$.  This is the hamiltonian of two
harmonic oscillators with unit mass and unit angular frequency, minus
a constant energy $M$.\footnote{In turn, the associate system can be
cast in covariant form as well.  Its extended phase space has
coordinates $(a,b,\tau)$ and its relativistic hamiltonian is
$H_{associate} = p_\tau + H$.}

The original system can be derived from the associate nonrelativistic
systems in two steps.  First, by restricting the motions to the ones
with total energy equal to zero (namely the ones in which the energy
$E=E_a+E_b$ of the two oscillators is $E=M$).  Second, by gauging away
the time evolution in $\tau$: that is, by restricting the physical
observables to the $\tau$ independent ones, namely to the sole
relations between $a$ and $b$.  This reduces by one the number of
degrees freedom.  Intuitively, we ``throw away the clock that measures
$\tau$".

\subsection{The Hamilton function}

As beautifully emphasized by Hamilton \cite{hamilton}, the solution of
any dynamical system is entirely coded in its Hamilton function.  This
is true for background independent systems as well.  Furthermore, the
Hamilton function plays a key role in relating the classical system to
the quantum system, and we shall use it extensively.  It is
essentially the classical limit of the quantum propagator.

The Hamilton function $S(a,b,a',b')$ is a function on ${\cal
C}\times{\cal C}$ that satisfies the Hamilton-Jacobi equation in both
sets of variables.  It is defined as the value of the action of the
physical motion that goes from $(a',b')$ to $(a,b)$.  If
$S(a,b,a',b')$ is known, all physical motions can be obtained just
taking derivatives, as follows.  Defining
\bea
p'_{a}(a,b,a',b')=\frac{\partial S(a,b,a',b')}{\partial a'},  
\hspace{1em}
%\nonumber\\ 
p'_{b}(a,b,a',b')=\frac{\partial S(a,b,a',b')}{\partial b'},   
\eea
the algebraic equations
\bea
p'_{a}(a,b,a',b')=p_{a}', 
\hspace{1em}
%\nonumber \\  
p'_{b}(a,b,a',b')=p_{b}'
\eea
are equivalent to (\ref{ellipses}) and give the physical motions. 
Here a motion is characterized by the ``initial data"
$(a',b',p'_{a},p'_{b})$ instead than by $A$ and $\phi$.  

Let us study the Hamilton functions of our system.  Given two points
$(a',b')$ and $(a,b)$ in $\cal C$, we first need to find the physical
motion that goes from one to the other.  That is, we need to find the
value of the two constants $A$ and $\phi$ of this motion.  To this
aim, it is convenient to parametrize the motion as in (\ref{eq:coord})
and search the value of the three constants $A$, $\phi$ and $\tau$
such that
\bea 
\left\{
\begin{array}{lclcl}
a' &=& a(\tau') &=& A\ \sin(\tau'+\phi), \\
b' &=& b(\tau') &=&  B\ \sin(\tau'); 
\end{array}\right.
\hspace{3em}
\left\{
\begin{array}{lclcl}
a &=& a(\tau'+\tau) &=& A\ \sin(\tau'+\tau +\phi), \\
b &=&  b(\tau'+\tau) &=& B\ \sin(\tau'+\tau ). 
\end{array}\right.
\eea
With some algebra we find 
\bea \label{eq:A1}
A^2 = \frac{a^2+a'^2-2aa' \cos \tau}{\sin^2 \tau}
\eea
and 
\bea 
\label{eq:M}
M=\frac{(a^2+b^2+a'^2+b'^2) -2(aa'+bb') \cos 
\tau}{\sin^2 \tau}
\eea
%
% where we have defined the two combinations of variables that will 
% appear repeatedly
% %
% \bea
% r & \equiv& r(a,b,a',b')\ = \ a^2+b^2+a'^2+b'^2, \label{R}\\
% t &  \equiv& t(a,b,a',b')\ = \  aa'+bb'. 
% \label{T}
% \eea 
%
The last equation can solved for $\tau$, giving
\bea 
\label{eq:tau}
\tau(a,b,a',b')=\arccos\frac{aa'+bb'\pm\sqrt{(aa'+bb')^2+M(M-a^2-b^2-a'^2-b'^2)}}{M}
\eea
Inserting this value in (\ref{eq:A1}) gives $A=A(a,b,a',b')$. 

The value of the Hamilton function is obtained by performing the
integration that defines the action (\ref{action}) of the motion that
goes from $(a',b')$ to $(a,b)$.  The result is
\begin{eqnarray}
\label{eq:Hfunction}
S(a,b,a',b')= S\Big(a,b,a',b';\ A(a,b,a',b')\Big)
\end{eqnarray}
where 
\begin{eqnarray}
S(a,b,a',b';\ A)=
S(a,b,A)-S(a',b',A).
\end{eqnarray}
Here $S(a,b,A)$ is given in (\ref{eq:HJfunction}) and $A(a,b,a',b')$
is the value of $A$ of the ellipse (\ref{ellipses}) that crosses
$(a,b)$ and $(a',b')$, determined by the equations (\ref{eq:A1}) and
(\ref{eq:tau}).

It easy to show that 
\bea 
\left.\frac{\partial S(a,b,a',b';\
A)}{\partial A}\right|_{A=A(a,b,a',b')} =0 
\eea 
We can also write the hamilton function in another form, that turns 
out
to be useful in the following.  Using (\ref{eq:coord}) and
(\ref{eq:M}) we can write
\begin{eqnarray}
S(a,b,a',b')=S(a,b,a',b';\ \tau(a,b,a',b')) 
\end{eqnarray}
where 
\bea 
\label{eq:action} 
S(a,b,a',b';\ \tau) = 
\frac{1}{2} \frac{(a^2+b^2+a'^2+b'^2) \cos \tau -2(aa'+bb')}{\sin
\tau} + M \tau .  
\eea 
This result is not surprising. The Hamilton function of a single 
harmonic oscillator is 
\bea 
\label{eq:actionHO} 
S(a,a'; \tau) = \frac{1}{2} \frac{(a^2+a'^2) \cos\tau -2aa'}{\sin 
\tau}; 
\eea 
equation (\ref{eq:action}) shows that the action of the relativistic
system is equal to the action of the associate system of two harmonic
oscillators $a(\tau)$ and $b(\tau)$ plus the constant $M$ term,
calculated for the time $\tau$ determined by the requirement that the
motion reproduces the constrained one of the covariant system. That 
is 
\begin{eqnarray}
S(a,b,a',b')=S(a,a'; \tau(a,b,a',b'))+ S(b,b'; \tau(a,b,a',b')) +
 M \tau(a,b,a',b').
\end{eqnarray}
Finally, it is easy to check that
\bea
\label{dopo}
\left.\frac{\partial S(a,b,a',b';\ \tau)}{\partial
\tau}\right|_{\tau=\tau(a,b,a',b')} =0 
\eea 
An equation that plays an important role below. 

\section{Quantum theory}

The classical system described above can be viewed as the classical
limit of a quantum system.  The quantum system is defined on the
kinematical Hilbert space ${\cal K}=L_{2}[R^2, da\, db]$ formed by the
wave functions $\psi(a,b)$.  The partial observables $a$ and $b$ act
as (self-adjoint) multiplicative operators.  The dynamics is defined
by the Wheeler-deWitt equation 
\bea 
\left(-\hbar^2\frac{\partial^2 }{\partial a^2}
-\hbar^2\frac{\partial^2}{\partial b^2} + a^2+b^2- 2M\right) 
\psi(a,b) = 0.
\label{WdW}
\eea
The operator in parenthesis is the Wheeler-deWitt operator $H$.  The
space of the solutions of this equation form the Hilbert space $\cal
H$ of the theory.  It is easy to find the general solution of this
equation.  The equation can be rewritten as
\bea 
H(a,b,p_a,p_b)\ \psi(a,b) = (H_{a}+H_{b}- M)\ \psi(a,b) = 0.
\label{WdW2}
\eea
where $H_{a}$ (and $H_{b}$) is the Shr\"odinger Hamiltonian operator
(with unit mass and unit frequency) in the $a$ (resp $b$) variable of 
a single harmonic oscillator. 
Since the eigenvalues of this operator are $E_{a}=\hbar (n_{a}+1/2)$
(and $E_{b}=\hbar (n_{b}+1/2)$) with non negative integer $n_{a}$ (and
$n_{b}$), in the basis of the energy eigenstates, $|n_a,n_b\rangle$, 
equation (\ref{WdW}) takes the form
\bea 
H(a,b,p_a,p_b)\ |n_a,n_b \rangle = (\hbar(n_{a}+n_{b}+1)- M)\ 
|n_a,n_b \rangle = 0.
\label{WdW21}
\eea
This equation has solution only if
\bea 
M=\hbar(n_{a}+n_{b}+1)
\eea
namely if $N=M/\hbar -1$ is a nonnegative integer. We shall assume so 
from 
now on.  The general non-normalized solution of (\ref{WdW}) is then 
easily 
\bea 
\psi(a,b)=\sum_{n_{a}+n_{b}+1=N} c_{n_{a},n_b} \psi_{n_{a}}(a)\
\psi_{n_{b}}(b)\ 
\eea
where $\psi_{n}$ is the normalized $n$-th eigenfunction of the
Harmonic oscillator, and the coefficients $c_{n_{a},n_b}$ satisfy 
$\sum_{n_{a}+n_{b}+1=N} |c_{n_{a},n_b}|^2 =1$.  The state space $\cal 
K$ is the Hilbert space of
the associate nonrelativistic system.  A basis in $\cal K$ is formed
by the states $|n_{a},n_{b}\rangle$ with $n_{a}$ $a$-quanta and
$n_{b}$ $b$-quanta:
\bea 
\langle a,b |n_{a},n_{b}\rangle = \psi_{n_{a}}(a)\
\psi_{n_{b}}(b)\ 
\eea
The physical Hilbert space $\cal H$ is the one spanned by the states with $N$
quanta.  It is convenient to write $j=N/2$ and to introduce the
quantum number $m=\frac{1}{2}(n_{a}-n_{b})$ that runs from $m=-j$ to
$m=j$. So that $\cal H$ is spanned by the $(2j+1)$ states 
\bea 
|m\rangle \equiv  |j-m,j+m\rangle.
\eea
That is 
\bea 
\langle a,b|m\rangle =  \psi_{j-m}(a)\ \psi_{j+m}(b)  \equiv 
\psi_{m}(a,b).
\eea
Explicitly, 
\bea 
\label{eq:statefunction3}
\psi_{m} (a,b)= 
\frac{1}{\scriptstyle{\sqrt{2^{2j}\hbar\pi 
(j\!+\!m)!(j\!-\!m)!}}}\ H_{j+m}(a/\sqrt{\hbar}) H_{j-m}(b/\sqrt{\hbar}) 
\ e^{-\frac{a^2+b^2}{2 \hbar}}
\eea
where $H_n(q)$ is the $n$-th Hermite polynomial.  (As pointed out by
Schwinger we can define angular momentum operators on the Hilbert
space of two oscillators; the total Hamiltonian is proportional to the
total angular momentum.  $\cal H$ is then the Hilbert space of the
irreducible spin-$j$ component of $\cal K$.  $m$ is the quantum number
associated to one of the components of the angular momentum:
\bea 
\label{eq:L}
 L_{z} &=& \frac{1}{4} (p_a^2 -p_b^2 + a^2 -b^2) \eea
see \cite{cr}.)  Alternatively, we can diagonalize the 
angular momentum 
\bea 
\label{eq:L2}
 L &=& \frac{1}{2} (b p_a -a p_b).
 \eea
Its eigenstates are easily written in polar coordinates $a=r\sin 
\varphi$,  $b=r\cos \varphi$ as 
\bea 
\label{eq:statefunction2}
\psi_{m'} (r,\varphi)= e^{\frac{i}{\hbar} m'\varphi}\psi_{m'}(r), 
\eea
where $m'$ is the eigenvalue of $L$ and $\psi_{m'}(r)$ is the 
(unique) solution of the radial equations 
with the correct energy
\bea
 \psi_{m'}(r)\ = \ e^{-r^2/2} \ r^{|m'|} \,_1F_1(-n,|m'|+1,r^2)
 \label{eq:hyper}
\eea
where $\,_1F_1(-n,|m'|+1,r^2)$ is the confluent hypergeometric 
function and $n=0,1,...,j$, $|m'|=0,1,..., 2j$, with the condition 
\bea
2j= 2n + |m'|.
\eea

$\cal H$ is a proper subspace of $\cal K$, therefore the scalar
product of $\cal K$ is well defined in $\cal H$.  As well known, in
several general relativistic systems the space of solutions of the
Wheeler-deWitt equation is a space of generalized states, namely a
subspace of a suitable completion of $\cal K$.  This is sometimes
described as serious conceptual difficulty, but it is not: there are
many equivalent techniques for deriving a scalar product of $\cal H$
from the scalar product of $\cal K$.  Here we are not concerned with
this technicality.

\subsection{The propagator}

Since $\cal H$ is a proper subspace of $\cal K$, there is an 
orthogonal 
projection operator 
\bea
P: {\cal K} \to {\cal H}. 
\eea
The relativistic propagator $K(a,b,a',b')$ is defined as the integral
kernel of $P$.  Easily, this projector is given by 
\bea
P = \sum_{n_{a}+n_{b}+1=N} |n_{a},n_{b}\rangle\langle n_{a},n_{b}|
\label{P}
\eea
Therefore 
\bea
K(a,b,a',b') = \sum_{n_{a}+n_{b}+1=N} \langle a,b|n_{a},n_{b}\rangle
\langle n_{a},n_{b}|a',b'\rangle. 
\eea
Explicitly, we can write this as 
\bea
K(a,b,a',b') &=& \sum_{m=-j}^j \Psi_{m}^* 
(a',b') \Psi_{m} (a,b) \\
&=& \sum_{m=-j} ^je^{-\frac{1}{2\hbar}(a^2+b^2+{a'}^2+{b'}^2)} \ 
\frac{H_{j+m}(a/\sqrt{\hbar}) H_{j-m}(b/\sqrt{\hbar})H_{j+m}(a'/\sqrt{\hbar}) 
H_{j-m}(b'/\sqrt{\hbar})}{{2^{2j}(j+m)!(j-m)! \hbar \pi}}
\eea
using the properties of the Hermite polynomials, this can be written 
for integer $j$
also as 
\bea
K(a,b,a',b') &=& \frac{1}{\pi}\  e^{-\frac{a^2+b^2+{a'}^2+{b'}^2}{2 \hbar}} 
\ 
\sum_{i=0} ^j \frac{(2(aa'+bb')/\hbar)^{2j-2i}}{(2j-2i)!}\ \  L_i^{2j-2i} 
\left( \frac{a^2+b^2+{a'}^2+{b'}^2}{\hbar} \right)
\eea
where $L_n^m$ is the $n$-th generalized Laguerre polynomial with
parameter $m$.
 
\subsection{Alternative expressions for the propagator}

The expression (\ref{P}) of the propagator can be rewritten as follows
\bea
P &=& \sum_{n_{a}+n_{b}+1=N} |n_{a},n_{b}\rangle\langle n_{a},n_{b}| 
%\nonumber\\ 
=
\sum_{n_{a},n_{b}} |n_{a},n_{b}\rangle\langle n_{a},n_{b}| \ 
\delta_{n_{a}+n_{a}+1-N} 
\nonumber\\ 
&=&\sum_{n_{a},n_{b}}  |n_{a},n_{b}\rangle\langle n_{a},n_{b}| 
\int_{0}^{2\pi}d\tau\  e^{-i(n_{a}+n_{b}+1-N)\tau}
%\nonumber\\ 
=\sum_{n_{a},n_{b}} |n_{a},n_{b}\rangle\langle n_{a},n_{b}| \
\int_{0}^{2\pi}d\tau\ e^{-\frac{i}{\hbar}H\tau} \nonumber\\
&=& \int_{0}^{2\pi}d\tau\  e^{-\frac{i}{\hbar}H\tau}.
\label{schw}
\eea
That is 
\be 
P\psi(a,b)=\int_{0}^{2\pi} d\tau\ \psi(a,b,\tau), 
\ee 
where $\psi(a,b,\tau)$ is the wave function of the associate system
that evolves in a time $\tau$ from the initial state
$\psi(a,b,0)=\psi(a,b)$.  The evolution of the associate system is
easy to determine.  We have
\be 
\psi(a,b,\tau)= e^{-\frac{i}{\hbar}H\tau} \psi(a,b,0)= \int da' \int 
db' 
K(a,a',\tau)K(b,b'\tau)e^{-\frac{i}{\hbar}M\tau}\ \psi(a,b,0)
\label{propin}
\ee 

where $K(a,a'\tau)$ is the well known propagator of a single Harmonic
oscillator
\be 
K(a,a',\tau)=\langle a |e^{-\frac{i}{\hbar}H_{a}\tau} |a'\rangle=
\frac{1}{\sqrt{2\pi i \hbar\sin}\tau} \exp\left(
\frac{i}{2\sin\tau}\left(\left[(a^2+a'^2)\cos\tau - 2 
aa'\right]\right)
\right), 
\label{propHO}
\ee 
and the term $e^{-\frac{i}{\hbar}M\tau}$ is present because of the
constant term in the hamiltonian.  Inserting (\ref{propHO}) in
(\ref{propin}) we find a simple integral expression for the propagator
\bea 
\label{eq:kk}
K(a,b,a',b') = \int_0 ^{2 \pi} d \tau \frac{1}{ 2 \pi i \hbar \sin 
\tau} \exp \left( \frac{i}{2 \hbar \sin \tau} \left[ 
(a^2+b^2+a'^2+b'^2) 
\cos \tau -2(a'a+b'b) \right] + \frac{i}{\hbar} M \tau 
\right)
\eea
that is
\bea 
\label{eq:kk27}
K(a,b,a',b') = \int_0 ^{2 \pi} d \tau \frac{1}{ 2 \pi i \hbar \sin 
\tau} \exp \left( \frac{i}{ \hbar } S(a,b,a',b'; \tau)
\right)
\eea
where $S(a,b,a',b'; \tau)$ is the hamilton function given by 
(\ref{eq:action}).

\subsection{Properties of the propagator}

Let us summarize some of the properties of the propagator.  The
propagator
\begin{itemize}
\item satisfies the Wheeler-deWitt equation (\ref{WdW}) in both sets
of variables.  \bea \left( - \hbar\frac{\partial^2}{\partial a^2}
-\hbar \frac{\partial^2}{\partial b^2} + a^2+b^2 - (2j+1)\hbar \right)
K(a,b,a',b') =0; \eea \item satisfies the composition law \bea \int
da'\int db'\ K(a,b,a',b') \ K(a',b',a'' ,b'') = K(a,b,a'',b''), \eea
this follows immediately from the fact that $P$ is a projector; \item
propagates a physical state from $(a',b')$ to $(a,b)$ in the sense
\bea \int da' db'\ K(a,b,a',b')\ \psi(a',b')= \psi(a,b); \eea \item
projects an arbitrary function $\psi(a,b)$ (not a solution of
(\ref{WdW})) to a solution of (\ref{WdW}); \item is gauge invariant,
in the sense that it is independent from the parameter $\tau$; \item
is real; \item is symmetric in the exchange $(a,b) \leftrightarrow
(a',b')$.
\end{itemize}

\section{Formal relations between quantum and classical theory}

From now we assume that $M\gg\hbar$ and we study the relation between
the classical and the quantum theory.

\subsection{The semiclassical limit of the propagator}
\label{semicllp}

We begin by relating the propagator $K(a,b,a',b')$ and the Hamilton
function $S(a,b,a',b')$.  A semiclassical approximation of the
propagator can be derived from the expression (\ref{eq:kk}).  In the
limit of small $\hbar$ we can evaluate the integral using a saddle
point approximation.  The result is
\bea 
\label{eq:kksc1}
K(a,b,a',b') = \sum_{i}\frac{1}{ 2 \pi i \hbar \sin 
\tau_{i}} \exp{\left(\frac{i}{2 \hbar \sin \tau_{i}} 
\left[ (a^2+b^2+a'^2+b'^2) 
\cos \tau_{i}  -2(a'a+b'b) \right] + \frac{i}{\hbar} M \tau_{i}
\right)}
\eea
where $\tau_{i}=\tau_{i}(a,b,a',b')$ are the values of $\tau$ for
which the exponential has an extremum.  That is, they are determined
by
\bea 
\label{eq:zeri}
\left.\frac{d}{d\tau}\left(\frac{i}{2 \hbar \sin \tau} 
\left[ (a^2+b^2+a'^2+b'^2) 
\cos \tau -2(a'a+b'b) \right] + \frac{i}{\hbar} M \tau
\right)\right|_{\tau=\tau_{i}}=0
\eea
But notice that the exponent is precisely $S(a,b,a',b',\tau)$, and
equation (\ref{eq:zeri}) is precisely equation (\ref{dopo}) that
determines the classical time $\tau$ along the classical trajectory
from $(a',b')$ to $(a',b')$.  Therefore the exponent in
(\ref{eq:kksc1}) is \emph{precisely} ($\frac{i}{\hbar}$) the Hamilton
function !

There is however an essential subtlety, that we have disregarded
during the classical discussion.  In general, equation (\ref{eq:zeri})
does not has one, but rather two solutions (a smooth periodic function
cannot have a single extremum).  Why two?  Because in general there
are two paths that connect two points $(a',b')$ and $(a,b)$: a shorter
one and a longer one obtained subtracting the shorter one from the
entire ellipses.  Indeed, it is clear that if $\tau$ is a solution, so
is $2\pi-\tau$.  Therefore the sum over $\tau_{i}$ contains two terms:
$\tau_{1}=\tau(a,b,a',b')$ where $\tau(a,b,a',b')$ is given by
(\ref{eq:tau}) and $\tau_{2}=2\pi-\tau(a,b,a',b')$.  Inserting this 
in  (\ref{eq:kksc}) we conclude that, to the first relevant order in 
$\hbar$, 
\bea 
K(a,b,a',b') = \frac{1}{2\pi i\hbar \sin \tau(a,b,a',b')}\ \left(
e^{-\frac{i}{\hbar}S(a,b,a',b')} - 
e^{+\frac{i}{\hbar}S(a,b,a',b')}\right),
\label{eq:kksc}
\eea
or 
\bea 
\label{eq:kksc11}
K(a,b,a',b') = \frac{1}{\pi \hbar \sin \tau(a,b,a',b')}\  \sin 
\left(\frac{1}{\hbar}{S(a,b,a',b')} \right).
\eea

\subsection{Quasi classical states}

Consider a coherent state of the associate nonrelativistic system. 
Using standard coherent state technology, the state determined by the
classical position $(a,b)$ and the classical momenta $(p_{a},p_{b})$
can be written as \bea
\left|a,b,p_{a},p_{b}\right\rangle = \sum_{n_a=0}^{\infty}  
\sum_{n_b=0} ^{\infty} \frac{\alpha_a ^{n_a} \alpha_b 
^{n_b}}{\sqrt{n_a! n_b!}} \exp 
\left(-\frac{|\alpha_a|^2+|\alpha_b|^2}{2} \right) \left| 
n_a,n_b \right\rangle
\eea
where $\alpha_{a}=a+ip_{a}$ and $\alpha_{b}=b+ip_{b}$.  This is a 
minimum spread wave packet centered in the position $(a,b)$ and with 
minimum spread momentum $(p_a,p_{b})$.  Recall that the coherent 
states of the harmonic oscillator have the property
\bea
\left|a,b,p_{a},p_{b},\tau\right\rangle \equiv 
e^{-\frac{i}{\hbar}H\tau}\left|a,b,p_{a},p_{b}\right\rangle = 
\left|a(\tau),b(\tau),p_{a}(\tau),p_{b}(\tau)\right\rangle 
\eea
where $(a(\tau),b(\tau),p_{a}(\tau),p_{b}(\tau))$ is the classical
motion determined by the initial conditions $a,b,p_{a},p_{b}$.  That
is, they follow classical trajectories exactly. 

Let us now project this state on $\cal H$. We obtain  
\bea
P\left|a,b,p_{a},p_{b} \right\rangle = \sum_{n_a+n_{b}+1=N}  
\frac{\alpha_a ^{n_a} \alpha_b 
^{n_b}}{\sqrt{n_a! n_b!}} \exp 
\left(-\frac{|\alpha_a|^2+|\alpha_b|^2}{2} \right) \left| 
n_a,n_b \right\rangle
\eea
On the other hand, we have also 
\bea 
\psi_{a,b,p_{a},p_{b}}(a',b')=\int_{0}^{2\pi} d\tau\
\psi_{a,b,p_{a},p_{b}}(a',b',\tau) 
\eea
where $\psi_{a,b,p_{a},p_{b}}(a,b,\tau)$ is the time evolution of the
coherent state of the associate system.  In the small $\hbar$ limit
the state $\psi(a,b,\tau)$ has essentially support only along a narrow
strip around the classical motion determined by the initial conditions
$(a,b,p_{a},p_{b})$.  Therefore the physical state
$\psi_{a,b,p_{a},p_{b}}(a',b')$ will have support on this same region,
namely on one of the ellipses (\ref{ellipses}).  The figure shows the
square of the amplitude of a coherent state $\psi(a,b)$ computed
numerically: it is picked around the classical solution.
\begin{figure}[h]
	\centering
	\includegraphics[width=7cm,angle=-90]{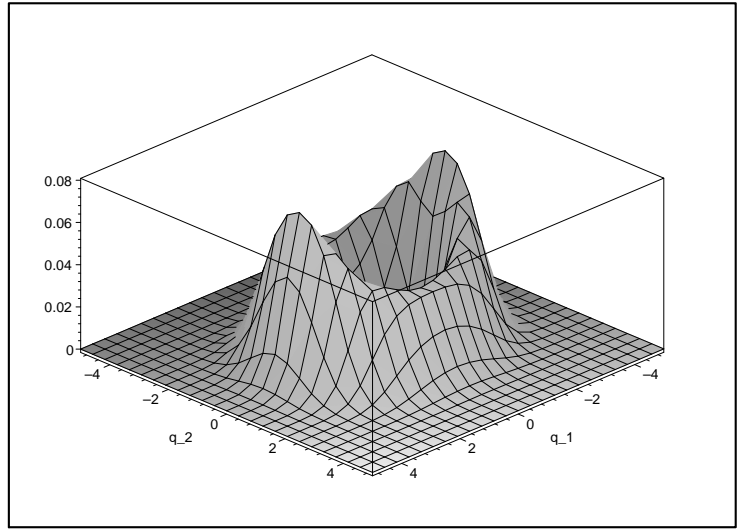}
	\caption{Support of a quasi classical state}
	\label{fig:SupportOfAQuasiClassicalState}
\end{figure}

\subsection{Forward and backward propagation}

In the classical theory, an ellipse defines a classical motion.  We
can assume that the versus of the ellipse, clockwise or anticlockwise,
has no physical significance.  In fact the equation (\ref{ellipses})
that fixes the motion does not contain any reference to the versus of
the ellipse.  However, in the quantum theory there are distinct
``clockwise and anticlockwise propagating" states.  To see this,
consider, instead of the coherent state defined in the previous
section, the one defined by the classical initial values
$(a,b,-p_{a},-p_{b})$.  In the associate system, the same ellipses is
followed in the opposite direction.  After integrating in $d\tau$, the
resulting physical coherent state $\psi_{a,b,-p_{a},-p_{b}}$ has the
same support as $\psi_{a,b,p_{a},p_{b}}$, but it is not the same state
!  A moment of reflection shows that it is its complex conjugate
\bea 
\psi_{a,b,-p_{a},-p_{b}}(a',b') = 
\psi^*_{a,b,p_{a},p_{b}}(a',b'). 
\eea
In fact, the physical Hilbert space can be split into the forward and 
backward propagating subspaces according to the sign of the 
eigenvalues of the operator $L$. 

A particularly simple semiclassical state, which illustrates well some
general features of these states, can be obtained as follows. 
Consider the classical motion formed by a circle, namely the case
$A=B=\sqrt{M}$.  The two corresponding semiclassical states are 
easily obtained. They are the states (\ref{eq:statefunction2}) 
with $m'=\pm 2j$, that maximize and minimize the angular momentum 
$L$. In the large $M/\hbar$ limit, the radial function 
become a narrow gaussian around the classical radius $\sqrt{M}$, as 
can be seen from (\ref{eq:hyper}). Indeed, when $m'=\pm 2j$, $n=0$, 
and since
\bea
\,_1F_1(-n,|m'|+1,r^2)=1
\eea
we have
\bea
 \psi_{\pm 2j}(r, \varphi)\ = \ e^{-r^2/2 +2j \ln r} \ e^{\pm 
\frac{i}{\hbar} 2j \varphi}. 
\eea
The maximum is in $r=\sqrt{2j} \approx \sqrt{M}$.

This observation sheds light on the fact that the propagator is real
and contains the two terms in the right hand side of (\ref{eq:kksc}). 
The propagator propagates both sectors of the Hilbert space -- the one
propagating clockwise, $\psi_{- 2j}(r, \varphi)$, and the one 
propagating anticlockwise, $\psi_{+ 2j}(r, \varphi)$.  Each of
the two terms of (\ref{eq:kksc}) propagates one of these sectors. 
Since the two are simply the complex conjugate of each other, the
propagator is real. 

On the other hand, there are quantum states that are in only one of
the two sectors.  They are ``clockwise propagating" states. 
Restricted to those states, the propagator can be written as a single
exponential of the Hamilton function.

Notice that the two terms of (\ref{eq:kksc}) can be interpreted in two
equivalent manners.  Either as related to the two branches of the
ellipses that connect $(a,b)$ and $(a',b')$.  Or as one term evolving
from $(a,b)$ to $(a',b')$ and other from $(a',b')$ to $(a,b)$.  The
math of section \ref{semicllp} can be interpreted either way.  Thus,
the second term of the propagator can be seen as the backward
propagating term.  We suspect that this term is therefore going to be
present irrespectively of whether the classical orbits are closed.

Recall that in quantum general relativity the three-geometry to
three-geometry transition amplitude has a surprising tendency of
turning out to be real.  For instance, the Ponzano-Regge transition
amplitudes are real.  Similarly, spin foam model tend to give real
transition amplitudes.  We suspect that the propagator of a covariant
theory is naturally defined as a real propagation --as in the example
studied here-- where there is no a priori distinction between forward
and backward propagation.  On the other hand, a quantum state may
contain one component only, and therefore be able to ``select" the
appropriate part of the propagator.  Form this point of view, the
attempt by Oriti and Livine to separate the two directions of
propagation in the spinfoam sums \cite{richard} can be seen as
attempts to separate locally the general relativistic analog of the
two terms of (\ref{eq:kksc}).

\section{Probabilistic interpretation}

The idea developed in \cite{partial,2,book} is that the variables of
the extended configuration space $\cal C$ can be interpreted as
partial observables, namely quantities that can be measured.  A point
in $\cal C$, called an ``event", represents a simultaneous measurement
of partial observables ($(a,b)$ in our case).  The classical theory
predicts which events can be measured: a classical state determines a
relation between partial observables and therefore a subset of events. 
The quantum theory assigns a probability amplitude for the measurement
of each event.  According to the general prescription given in
\cite{2}, the probability of observing an event in a small region
$\cal R$ of $\cal C$ if an event has been observed in a small region
${\cal R}'$ is
\be
      P_{{\cal R}{\cal R}'}=\frac{|K_{{\cal R}{\cal R}'}|^2}
      {{K_{{\cal R}{\cal R}}\ K_{{\cal R}'{\cal R}'}}}      
      \label{pro1}
\ee
where 
\be
      K_{{\cal R}{\cal R}'}=\int_{\cal R}dadb \int_{{\cal 
      R}'}da'db' \ \ K(a,b,a',b')  .     
\ee
The propagator $K$ contains then the full physically relevant
information about the quantum theory.  Alternatively, given a physical
state $\psi$, the probability for detecting an event around a point
$(a,b)$ in a small region $\cal R$ of volume $V$, where $\psi(a,b)$
can be taken to constant, is
\be
\label{pro}
      P_{{\cal R}}\ =\  \frac{\left| \int_{R} \psi(a,b)\ da \, db \right| ^2} 
      {{K_{{\cal R}{\cal R}}}}
     \ =\   \frac{V^2} 
      {{K_{{\cal R}{\cal R}}}}\ \  |\psi(a,b)| ^2. 
\ee

To test the viability of this interpretation in the context of our
system, we must relate it to the standard interpretation of quantum
mechanics.  According to the standard interpretation of the wave
function $\psi(x,t)$, the modulus square of the wave function \emph{at
fixed time} is to be interpreted as the probability density \emph{in
space} for the position of the particle.  A probability is therefore
given by
\be
       dP = |\psi(x,t)|^2 dx.
\ee
Can we recover this expression from (\ref{pro}) in a limit in which
our system behaves as a nonrelativistic system where, say, $b$ is
taken as the independent variable (as $t$) and $a$ as a dynamical
variable (as $x$) varying in time?

For this, we must look for a regime where the fact that the evolution
``comes back" in the time $b$ is negligible.  Consider a region $\cal
R$ such that $a^2,b^2 \ll M$.  As in \cite{2}, imagine we have a
detector active for very short ``time" $b$.  That is, consider a small
detection region $\Delta b \ll (\Delta a)^2 \ll 1$.  Let us then
calculate the detection probability using (\ref{pro}).  In the
region, $\tau(a,b,a',b') \ll 1$.  Assume also that $M \gg \hbar$ and
we can use the saddle point approximation. The propagator takes
the form
\bea 
K(a,b,a',b')=
\frac{1}{\pi\hbar \tau} \sin \left(\frac{(a-a')^2+(b-b')^2}{2 \hbar\tau}+ 
\frac{M}{\hbar} \tau
\right) 
\eea
Using the expression of the parameter $\tau$ from
(\ref{eq:M}) 
\bea 
K(a,b,a',b') = \frac{\sqrt{M}}{\pi\hbar} \frac{\sin
\left( \frac{3}{2\hbar} \sqrt{M}\sqrt{(a-a')^2+(b-b')^2}
\right)}{\sqrt{(a-a')^2+(b-b')^2}} 
\eea 
Since $M \gg \hbar$, this gives 
\bea K(a,b,a',b') &\approx&
\frac{3\sqrt{M}}{2\pi\sqrt{\hbar}} \  \delta \left(
\sqrt{(a-a')^2+(b-b')^2}/\hbar \right) 
\eea Performing the integration in the region $\cal R$, and using our
conditions on the shape of $\cal R$ (which are not symmetric in $a$
and $b$) we obtain
\be K_{{\cal R}{\cal R}}=\int_{\cal
R}dadb \int_{{\cal R}}da'db' \ \ K(a,b,a',b') \approx
\frac{3\sqrt{M}}{2\pi\hbar^2} \ \Delta a (\Delta b)^2 
\ee 
so \bea \label{prob} P_{{\cal R}}= \frac{\left| \int_{R} \psi da \, db
\right| ^2} {{K_{{\cal R}{\cal R}}}} \approx \frac{|\psi(a,b) |^2
(\Delta a)^2(\Delta b)^2}{\Delta a (\Delta b)^2} = |\psi(a,b) |^2
\Delta a \eea where we have absorbed a proportionality constant in the
normalization of the wave function.  The key result is that the
probability is independent from $\Delta b$ and proportional to $\Delta
a$.  The first fact allows us to use the notion of measurement
\emph{istantaneous} in $b$, because the ``time" $b$ needed by the
apparatus to perform the measurmenet has no effect on the result.  The
second implies that $ |\psi(a,b) |^2 $ can be interpreted as a
probability density in $a$ at fixed $b$.  Therefore we have recovered
the standard probabilistic interpretation of the wave function in
quantum mechanics: $|\psi|^2$ is the probability density in space, at
fixed time.

\section{Conclusion}

We have studied a simple dynamical system that reproduces some key
aspects of the background independence of general relativity.  We have
used this system to illustrate some general features of the structure
of general covariant dynamical systems.  

According to the point of view we have taken here, the dynamics of a
system with $n$ degrees of freedom does not describe the evolution in
time of $n$ observables.  Rather, it describes the correlations
between $n+1$ partial observables.  The space of the partial
observables is the extended configuration space $\cal C$, and the
dynamics is governed by a (vanishing) relativistic hamiltonian $H$ on
$T^*{\cal C}$.  In the quantum theory, the kinematical Hilbert space
$\cal K$ expresses all potential outcomes of measuments of partial
observables.  Dynamics is a restriction on these states and expresses
the existence of correlations among measurements of partial
observables.  Such restriction of the states in $\cal K$ is given by
the Wheeler-deWitt equation $H\psi=0$.  Given a physical state, the
probability that the system is detected in a small region of $\cal C$
is governed by (\ref{pro}).  (A discussion of the precise meaning of
probability in this context is in \cite{2}.)  Here we have shown that
in our simple system this probability prescription reduces to the
conventional one in the appropriate regime.

This definition of probability yields positive probabilities, unlikely
the definition of probability as a flux of a current (see Appendix B),
often used in quantum cosmology \cite{halliwell}.  It is reasonable to
expect the two definitions to agree for states that propagate in only
one direction.  We also expect it to agree with the probabaility
computed with histories techniques studied in \cite{halliwell} and we
think that the precise relation between these two points of view
deserves to be studied.

The kernel of the projector from the kinematical to the physical state
space is the propagator, which codes the dynamical content of the
theory and can be taken as the basis of the probabilistic
interpretation.  We have studied the propagator of our model in
detail.  We have shown that in the semiclassical limit it has a simple
relation with the Hamilton function of the classical theory, but this
relation is not a simple exponential, as one might have expexted. 
Instead, the propagator is real.  It is the sum of two exponential
terms complex conjugate to each other, that propagate backward and
forward, respectively, along the motions.  Accordingly, the physical
Hilbert space splits between forward and backward propagating states. 
We expect this structure to be the same in quantum general relativity.

\appendix
\section{Appendix: Approximate Schr\"odinger picture}
If we want to describe just the local evolution, we can cast the
problem in a Schr\"odinger form by choosing an ``internal time"
variable as a function on the phase-space, before quantizing.  The
starting point for this is to find a function $T(a,b,p_a,p_{b})$ that
satisfies
\be
\left\{T,H \right\}=1, 
\label{TH}
\ee
We have then $dT/d \tau =1$, here $\tau$ is the evolution parameter. 
We chose 
\be
T(b,p_b) = \arctan \left( \frac{b}{p_b} \right), 
\ee
that satisfies (\ref{TH}). 
Next, we search for a canonnical transformation from the canonical 
pair $b,p_{b}$ to a canonical pair $T,P_{T}$. This is given by 
\bea
b &=& \pm \sqrt{2P_{T}} \sin T, \\
p_b &=& \pm \sqrt{2P_{T}} \cos T. 
\eea
A generating function for this canonical 
transformation is
\be
       S(b,T) =  \frac{b^2}{2\tan T}.
\ee
Indeed, we have 
\be 
        P= -\frac{ \partial S}{ \partial Q} 
\ee 
and 
\be 
      p_{b}= -\frac{ \partial S}{ \partial b}. 
\ee 
The Hamiltonian in the canonical variables $(a,T,p_{a},P_{T})$ is then 
\be
     H= P_{T} + H_{eff} 
\ee 
where 
\be
       H_{eff}= \frac{1}{2} (p_a^2 + a^2) -M
\ee
from which we obtain the Schr\"odinger
equation
\be 
    -i \hbar \frac{ \partial \psi}{ \partial T} = H_{eff} \psi 
\ee
It is important to notice that the range of $T$ is contained in the
interval $[-\frac{\pi}{2}, \frac{\pi}{2}]$, therefore this change of
variables cannto be used around the points $b=\pm \frac{\pi}{2}$,
where the ``time" $b$ ``comes back".  The propagator is, up to a
phase, the one of the standard harmonic oscillator.  
\be
\label{eq:kk2} K(a,T ,a',0) = \frac{1}{ \sqrt{2 \pi i \hbar \sin T}}
\exp \left( \frac{i}{2 \hbar \sin T} \left[ (a^2+a'^2) \cos T -2a'a \right]
+ \frac{i}{\hbar} M T \right). 
\ee

\section{WKB}

We recall here the interpretation of the semiclassical 
approximation of a quantum cosmological model in terms of the 
conserved current of the WKB approximation. 

Let us search approximate solutions of the Wheeler-deWitt equation 
(\ref{WdW}) in the form 
\bea 
\label{eq:WKB}
\psi(a,b) = A(a,b)\  e^{ \frac{i}{\hbar} S(a,b)}, 
\eea
where $A(a,b)$ is a function that varies slowly, in a sense that we
specify in a moment.  Inserting this function in (\ref{WdW}) we obtain
for the real part 
\bea 
\frac{\partial^2 A}{\partial^2 a} +
\frac{\partial^2 A}{\partial^2 b} -\frac{A}{\hbar ^2} \left( \left(
\frac{\partial S}{\partial a}\right)^2 +\left( \frac{\partial
S}{\partial b}\right)^2 \right) = \frac{-2(2j+1) +a^2+b^2}{\hbar ^2}
A. 
\eea 
If $A(a,b)$ varies slowly in the sense 
\bea
\frac{\frac{\partial^2 A}{\partial^2 a} + \frac{\partial^2
A}{\partial^2 b}}{A} \ll \frac{-2(2j+1) +a^2+b^2}{\hbar ^2}, 
\eea
then $S(a,b)$ must satisfy the Hamilton-Jacobi equation 
\bea
\frac{1}{2} \left( \frac{\partial S}{\partial a}\right)^2 +
\frac{1}{2} \left( \frac{\partial S}{\partial b}\right)^2 +
\frac{a^2}{2} + \frac{b^2}{2} -(2j+1) =0 
\eea 
For the imaginary part we have
\bea
2 \frac{\partial A}{\partial a} \frac{ \partial S}{\partial 
a} +2 \frac{\partial A}{\partial b} \frac{ \partial 
S}{\partial b}  +A\left( \frac{\partial^2 S}{\partial^2 a} 
+\frac{\partial^2 S}{\partial^2 b} \right)=0
\eea
which can be written as
\bea \label{eqwkb}
\frac{\partial }{\partial a} \left(A^2 \frac{ \partial 
S}{\partial a} \right)+\frac{\partial }{\partial b} 
\left(A^2 \frac{ \partial S}{\partial b} \right)=0
\eea
and interpreted as a continuity equation $\partial_a j^a+\partial_b 
j^b=0$ for the current $\vec j=(j^a,j^b)$
\bea 
\vec j&=& A^2\ \vec\nabla S. 
\eea
where 
\bea 
\vec\nabla S = \left(\frac{\partial S}{\partial
a},\frac{\partial S}{\partial b} \right) 
\eea 
The WKB approximation is a widely used technique in quantum
cosmological models.  The central question is how to extract physical
predictions from a WKB solution.  The Wheeler-DeWitt equation is
typically a second-order equation like the Klein-Gordon equation, and
the associated current 
\bea 
    \vec J= i (\psi \vec \nabla \psi^* - \psi^* \vec \nabla \psi)
\eea
is conserved: $div\vec J =0$.  This current $J$ is intended to provide
the interpretation of the WKB solution: the flux of $\vec J$ across a
surface $\Sigma$ defines the probability that the set of classical
trajectories with momentum $\vec p= \vec \nabla S$ corresponding to
the wave function intersect the surface $\Sigma$.  In general, this
current produces negative probabilities.  In particular, we have a
full basis of real solutions where $\vec J=0$.

\section{Path integral representation of the propagator}

Finally, we illustrate a path integral derivation of the propagator. 
The action of the relativistic system can be formulated on an extended
configuration space of the coordinates $a,b$ and a lagrangian
multiplier $N$ that implements the constraint (\ref{Huz}) \bea
S[a,b,N]= \int_0 ^1 d\tau \left( p_a \dot{a} + p_b \dot{b} -
N\frac{1}{2}(p_a^2+p_b^2+a^2+b^2-2M) \right) \eea where a dot denotes
the derivative with respect to $\tau$.  The variation of the action
with respect to the variables $a(\tau), p_a(\tau), b(\tau), p_b(\tau)$
and $N(\tau)$ gives the canonical equations of motion and the
constraint equation.  The action is invariant under reparametrization
of the parameter $\tau$ labeling the motion.  This invariance implies
a relation between the velocities and the momenta, obtained varying
the action with respect to $p_a$ and $p_b$:$ \dot{a}= N p_a$ and
$\dot{b}=N p_b$.  The action can be put in the form \bea S[a,b,N]=
\frac{1}{2} \int_0 ^1 d\tau\ \left( \frac{\dot{a}^2 + \dot{b}^2}{N}
-N(a^2+b^2-2M) \right) \eea which has the same structure as the ADM
action of general relativity.  The path integral representation of the
propagator of the quantized theory has the form \bea K(a',b',a,b)=
\int dN \int_a ^{a'} \mathcal{D} a \int_b ^{b'} \mathcal{D} b \;
e^{iS[a,b,N]} \eea In order to sum over only inequivalent trajectories
we fix the gauge in the action choosing $N(\tau)= constant$.  After
gauge-fixing and after rescaling the parameter $\tau$ as
$\tilde{\tau}= \tau N$, the propagator can be written \bea
K(a',b',a,b)= \int dN \int_a ^{a'} \mathcal{D} a \int_b ^{b'}
\mathcal{D} b \exp \left[ i \int_0 ^N d \tilde{\tau} \frac{1}{2}
\left( {\dot{a}^2 + \dot{b}^2} -a^2-b^2+2M \right) \right] \nonumber
\eea The range of integration of the lagrangian multiplier will be
fixed by the symmetry property of the integrand.  the explicit
expression of the propagator results to be
 
 \bea
 K(a',b',a,b)  &=& \int dN \mathcal{D} a \mathcal{D} b \exp \left[ i 
\int_0 ^N d\tilde{\tau} \frac{1}{2} \left( \dot{a}^2 + \dot{b}^2 
-a^2-b^2 \right) +iMN \right] \nonumber \\
  &=& \int dN \left(\int \mathcal{D} a  \exp \left[i \int_0 ^N 
d\tilde{\tau} \frac{1}{2} \left( \dot{a}^2  -a^2 \right) \right] 
\right) \left(\int \mathcal{D} b \exp \left[ i \int_0 ^N 
d\tilde{\tau} \frac{1}{2} \left( \dot{b}^2 -b^2 \right)\right]\right) 
e^{iMN} \nonumber
  \eea
The expressions in parenthesis have the form of the path integral for 
the propagator of a free one-dimensional harmonic oscillator over the 
time $N$ in the coordinates $a$ and $b$ respectively:  
  \bea
  K(a',b',a,b) = \int dN K(a',a,N) K(b',b,N) e^{iMN}
 \eea
 The propagator of a free  harmonic oscillator is a periodic function 
of the time:
 \bea
 K(a',a,N+2 \pi)= K(a',a,N)
 \eea
 Consequently we are free to choose the interval $[0,2 \pi]$ as the 
range of integration for $N$
 \bea
 K= \int_0 ^{2 \pi} dN K(a',a,N) K(b',b,N) e^{iMN}.
 \eea
 This representation of the propagator is exactly the expression 
(\ref{eq:kk}).

\end{document}